

\input harvmac

\overfullrule=0pt



\def\bs{\bigskip}
\def\no{\noindent}
\def\hb{\hfill\break}
\def\qq{\qquad}
\def\bl{\bigl}
\def\br{\bigr}

\def\IR{\relax{\rm I\kern-.18em R}}

\def\np {  Nucl. Phys. }


\def\r{\rho}
\def\a{\alpha}

\def\g{\gamma}
\def\G{\Gamma}
\def\d{\delta}

\def\m{\mu}
\def\n{\nu}

\def\l{\lambda}
\def\L{\Lambda}
\def\s{\sigma}

\def\IR{\relax{\rm I\kern-.18em R}}

\def\pap{\partial_+}
\def\pam{\partial_-}
\def\papm{\partial_{\pm}}

\def\pat{\partial_{\tau}}



hep-th/9305074 \hfill {USC-93/HEP-S1}
\rightline{May 1993}
\bs

\centerline  {\bf EFFECTIVE ACTION AND EXACT GEOMETRY IN  }
\centerline   {   {\bf CHIRAL GAUGED WZW MODELS}
                  {  \footnote{$^\dagger$}
 {Research supported in part by DOE, under Grant No. DE-FG03-84ER-40168. } }
               }

\vskip 1.00 true cm

\centerline  { { \bf Konstadinos Sfetsos }
              \footnote {$^{*}$}
 {Address after September of 93: Institute for Theoretical Physics,
  University of Utrecht, The Netherlands.  }    }

\bigskip

\centerline {Physics Department}
\centerline {University of Southern California}
\centerline {Los Angeles, CA 90089-0484, USA}


\vskip 1.50 true cm

\centerline{ABSTRACT}

\vskip .3 true cm

Following recent work on the effective quantum action of gauged WZW models,
we suggest such an action for {\it chiral} gauged WZW models which
in many respects differ from the usual gauged WZW models.
Using the effective action we compute the conformally exact expressions
for the metric, the antisymmetric tensor, and the dilaton fields
in the $\s$-model arising from a general {\it chiral } gauged WZW model.
We also obtain the general solution of the geodesic equations in the
exact geometry.
Finally we consider in some detail a three dimensional model
which has certain similarities with the three dimensional black string model.

\vfill\eject


\lref\BSthree{I. Bars and K. Sfetsos, Mod. Phys. Lett. {\bf A7} (1992) 1091.}

\lref\BShet{I. Bars and K. Sfetsos, Phys. Lett. {\bf 277B} (1992) 269.}

\lref\BSglo{I. Bars and K. Sfetsos, Phys. Rev. {\bf D46} (1992) 4495.}

\lref\BSexa{I. Bars and K. Sfetsos, Phys. Rev. {\bf D46} (1992) 4510.}

\lref\SFET{K. Sfetsos, Nucl. Phys. {\bf B389} (1993) 424.}

\lref\BSslsu{I. Bars and K. Sfetsos, Phys. Lett. {\bf 301B} (1993) 183.}

\lref\BSeaction{I. Bars and K. Sfetsos ``Exact Effective Action and Space-time
Geometry in Gauged WZW Models'', USC-93/HEP-B1 (hep-th/9301047),
to appear in Phys. Rev. {\bf D} (1993).}

\lref\BN{ I. Bars and D. Nemeschansky, Nucl. Phys. {\bf B348} (1991) 89.}

\lref\WIT{E. Witten, Phys. Rev. {\bf D44} (1991) 314.}

 \lref\IBhet{ I. Bars, Nucl. Phys. {\bf B334} (1990) 125. }

 \lref\IBCS{ I. Bars, {\it Curved Space-time Strings and Black Holes},
in Proc.
 {\it XX$^{th}$ Int. Conf. on Diff. Geometrical Methods in Physics}, eds. S.
 Catto and A. Rocha, Vol. 2, p. 695, (World Scientific, 1992).}

 \lref\CRE{M. Crescimanno, Mod. Phys. Lett. {\bf A7} (1992) 489.}

\lref\clapara{K. Bardakci, M. Crescimanno and E. Rabinovici,
Nucl. Phys. {\bf B344} (1990) 344.}

\lref\MSW{G. Mandal, A. Sengupta and S. Wadia,
Mod. Phys. Lett. {\bf A6} (1991) 1685.}

 \lref\HOHO{J. B. Horne and G. T. Horowitz, Nucl. Phys. {\bf B368} (1992) 444.}

 \lref\FRA{E. S. Fradkin and V. Ya. Linetsky, Phys. Lett. {\bf 277B}
          (1992) 73.}

 \lref\ISH{N. Ishibashi, M. Li, and A. R. Steif,
         Phys. Rev. Lett. {\bf 67} (1991) 3336.}

 \lref\HOR{P. Horava, Phys. Lett. {\bf 278B} (1992) 101.}

 \lref\RAI{E. Raiten, ``Perturbations of a Stringy Black Hole'',
         Fermilab-Pub 91-338-T.}

 \lref\GER{D. Gershon, ``Exact Solutions of Four-Dimensional Black Holes in
         String Theory'', TAUP-1937-91.}

 \lref \GIN {P. Ginsparg and F. Quevedo,  Nucl. Phys. {\bf B385} (1992) 527. }

 \lref\HOHOS{ J. H. Horne, G. T. Horowitz and A. R. Steif, Phys. Rev. Lett.
 {\bf 68} (1991) 568.}

 \lref\groups{
 M. Crescimanno. Mod. Phys. Lett. {\bf A7} (1992) 489. \hb
 E. S. Fradkin and V. Ya. Linetsky, Phys. Lett. {\bf 277B} (1992) 73. \hb
 P. Horava, Phys. Lett. {\bf 278B} (1992) 101.\hb
 E. Raiten, ``Perturbations of a Stringy Black Hole'',
         Fermilab-Pub 91-338-T.\hb
 D. Gershon, ``Exact Solutions of Four-Dimensional Black Holes in
         String Theory'', TAUP-1937-91.}

\lref\NAWIT{C. Nappi and E. Witten, Phys. Lett. {\bf 293B} (1992) 309.}

\lref\FRATSE{E. S. Fradkin and A. A. Tseytlin,
Phys. Lett. {\bf 158B} (1985) 316.}

\lref\CALLAN{ C. G. Callan, D. Friedan, E. J. Martinec and M. Perry,
Nucl. Phys. {\bf B262} (1985) 593.}

\lref\DB{L. Dixon, J. Lykken and M. Peskin, Nucl. Phys.
{\bf B325} (1989) 325.}

\lref\IB{I. Bars, Nucl. Phys. {\bf B334} (1990) 125.}

\lref\BUSCHER{T. Buscher, Phys. Lett. {\bf 159B} (1985) 127.\hb
T. Buscher, Phys. Lett. {\bf 194B} (1987) 59.\hb
T. Buscher, Phys. Lett. {\bf 201B} (1988) 466.}

\lref\Dual{M. Rocek and E. Verlinde, Nucl. Phys. {\bf B373} (1992) 630.\hb
A. Giveon and M. Rocek, Nucl. Phys. {\bf B380} (1992) 128. }

\lref\nadual{X. C. de la Ossa and F. Quevedo, ``Duality Symmetries from
Non-abelian Isometries in String Theory'', NEIP-92-004.}

\lref\SENrev{A. Sen, ``Black Holes and Solitons in String Theory'',
TIFR-TH-92-57.}

\lref\TSEd{A. A. Tseytlin, Mod. Phys. Lett. {\bf A6} (1991) 1721.}

\lref\TSESC{A. S. Schwarz and A. A. Tseytlin, ``Dilaton shift under duality
and torsion of elliptic complex'', IMPERIAL/TP/92-93/01. }

\lref\Dualone{K. Meissner and G. Veneziano,
Phys. Lett. {\bf B267} (1991) 33. \hb
K. Meissner and G. Veneziano, Mod. Phys. Lett. {\bf A6} (1991) 3397. \hb
M. Gasperini and G. Veneziano, Phys. Lett. {\bf 277B} (1992) 256. \hb
M. Gasperini, J. Maharana and G. Veneziano, Phys. Lett. {\bf 296B} (1992) 51.}

\lref\Dualtwo{A. Sen,
Phys. Lett. {\bf B271} (1991) 295;\ ibid. {\bf B274} (1992) 34. \hb
A. Sen, Phys. Rev. Lett. {\bf 69} (1992) 1006. \hb
S. Hassan and A. Sen, Nucl. Phys. {\bf B375} (1992) 103. \hb
J. Maharana and J. H. Schwarz, Nucl. Phys. {\bf B390} (1993) 3.}

\lref\KIRd{E. Kiritsis, ``Exact Duality Symmetries in CFT and String Theory'',
LPTENS-92-29; CERN-TH-6797-93.}

\lref\GIPA{A. Giveon and A. Pasquinucci, ``On cosmological string backgrounds
with toroidal isometries'', IASSNS-HEP-92/55, August 1992.}

\lref\KASU{Y. Kazama and H. Suzuki, Nucl. Phys. {\bf B234} (1989) 232. \hb
Y. Kazama and H. Suzuki Phys. Lett. {\bf 216B} (1989) 112.}

\lref\WITanom{E. Witten, Comm. Math. Phys. {\bf 144} (1992) 189.}

\lref\WITnm{E. Witten, Nucl. Phys. {\bf B371} (1992) 191.}

\lref\IBhetero{I. Bars, Phys. Lett. {\bf 293B} (1992) 315.}

\lref\IBerice{I. Bars, {\it Superstrings on Curved Space-times}, Lecture
delivered at the Int. workshop on {\it String Quantum Gravity and Physics
at the Planck Scale}, Erice, Italy, June 1992.}

\lref\DVV{R. Dijkgraaf, E. Verlinde and H. Verlinde, Nucl. Phys. {\bf B371}
(1992) 269.}

\lref\TSEY{A. A. Tseytlin, Phys. Lett. {\bf 268B} (1991) 175.}

\lref\JJP{I. Jack, D. R. T. Jones and J. Panvel,
          Nucl. Phys. {\bf B393} (1993) 95.}

\lref\BST { I. Bars, K. Sfetsos and A. A. Tseytlin, unpublished. }

\lref\TSEYT{ A. A. Tseytlin, ``Effective Action in Gauged WZW Models
and Exact String Solutions", Imperial/TP/92-93/10.}

\lref\TSEYTt{A. A. Tseytlin, ``Conformal Sigma Models corresponding to Gauged
WZW Models'', CERN-TH.6804/93.}

 \lref\SHIF { M. A. Shifman, Nucl. Phys. {\bf B352} (1991) 87.}
\lref\SHIFM { H. Leutwyler and M. A. Shifman, Int. J. Mod. Phys. {\bf
A7} (1992) 795. }

\lref\POLWIG { A. M. Polyakov and P. B. Wiegman, Phys.
Lett. {\bf 141B} (1984) 223.  }

\lref\BCR{K. Bardakci, M. Crescimanno
and E. Rabinovici, Nucl. Phys. {\bf B344} (1990) 344. }

\lref\Wwzw{E. Witten, Commun. Math. Phys. {\bf 92} (1984) 455.}

\lref\GKO{P. Goddard, A. Kent and D. Olive, Phys. Lett. {\bf 152B} (1985) 88.}

\lref\Toda{A. N. Leznov and M. V. Saveliev, Lett. Math. Phys. {\bf 3} (1979)
489. \hb A. N. Leznov and M. V. Saveliev, Comm. Math. Phys. {\bf 74}
(1980) 111.}

\lref\GToda{J. Balog, L. Feh\'er, L. O'Raifeartaigh, P. Forg\'acs and A. Wipf,
Ann. Phys. (New York) {\bf 203} (1990) 76.; Phys. Lett. {\bf 244B}
(1990) 435.}

\lref\GWZW{ E. Witten, \np {\bf B223} (1983) 422. \hb
K. Bardakci, E. Rabinovici and B. S\"aring, Nucl. Phys. {\bf B299}
(1988) 157. \hb K. Gawedzki and A. Kupiainen, Phys. Lett. {\bf 215B}
(1988) 119. \hb K. Gawedzki and A. Kupiainen, Nucl. Phys. {\bf B320}
(1989) 625. }

\lref\SCH{ D. Karabali, Q-Han Park, H. J. Schnitzer and Z. Yang,
                   Phys. Lett. {\bf B216} (1989) 307. \hb D. Karabali
and H. J. Schnitzer, Nucl. Phys. {\bf B329} (1990) 649. }

 \lref\KIR{E. Kiritsis, Mod. Phys. Lett. {\bf A6} (1991) 2871. }

\lref\BIR{N. D. Birrell and P. C. W. Davies,
{\it Quantum Fields in Curved Space}, Cambridge University Press.}

\lref\WYB{B. G. Wybourn, {\it Classical Groups for Physicists }
(John Wiley \& sons, 1974).}

\lref\SANTA{R. Guven, Phys. Lett. {\bf 191B} (1987) 275.\hb
D. Amati and C. Klimcik, Phys. Lett. {\bf 219B} (1989) 443.\hb
G. T. Horowitz and A. R. Steif, Phys. Rev. Lett. {\bf 64} (1990) 260.\hb
G. T. Horowitz and A. R. Steif, Phys. Rev. {\bf D42} (1990) 1950.}

\lref\SANT{J. H. Horne, G. T. Horowitz and A. R. Steif,
Phys. Rev. Lett. {\bf 68} (1991) 568.}

\lref\PRE{J. Prescill, P. Schwarz, A. Shapere, S. Trivedi and F. Wilczek,
Mod. Phys Lett. {\bf A6} (1991) 2353.\hb
C. Holzhey and F. Wilczek, Nucl. Phys. {\bf B380} (1992) 447.}

\lref\HAWK{J. B. Hartle and S. W. Hawking Phys. Rev. {\bf D13} (1976) 2188.\hb
S. W. Hawking, Phys. Rev. {\bf D18} (1978) 1747.}

\lref\HAWKI{S. W. Hawking, Comm. Math. Phys. {\bf 43} (1975) 199.}

\lref\HAWKII{S. W. Hawking, Phys. Rev. {\bf D14} (1976) 2460.}

\lref\euclidean{S. Elitzur, A. Forge and E. Rabinovici,
Nucl. Phys. {\bf B359} (1991) 581. }

\lref\ITZ{C. Itzykson and J. Zuber, {\it Quantum Field Theory},
McGraw Hill (1980). }

\lref\kacrev{P. Goddard and D. Olive, Journal of Mod. Phys. {\bf A} Vol. 1,
No. 2 (1986) 303.}

\lref\BBS{F. A. Bais, P. Bouwknegt and M. Surridge, Nucl. Phys. {\bf B304}
(1988) 348.}

\lref\nonl{A. Polyakov, {\it Fields, Strings and Critical Phenomena}, Proc. of
Les Houses 1988, eds. E. Brezin and J. Zinn-Justin North-Holland, 1990.\hb
Al. B. Zamolodchikov, preprint ITEP 87-89. \hb
K. Schoutens, A. Sevrin and P. van Nieuwenhuizen, Proc. of the Stony Brook
Conference {\it Strings and Symmetries 1991}, World Scientific,
Singapore, 1992. \hb
J. de Boer and J. Goeree, ``The Effective Action of $W_3$ Gravity to all
\hb orders'', THU-92/33.}

\lref\HOrev{G. T. Horowitz, {\it The Dark Side of String Theory:
Black Holes and Black Strings}, Proc. of the 1992 Trieste Spring School on
String Theory and Quantum Gravity.}

\lref\HSrev{J. Harvey and A. Strominger, {\it Quantum Aspects of Black
Holes}, Proc. of the 1992 Trieste Spring School on
String Theory and Quantum Gravity.}

\lref\GM{G. Gibbons, Nucl. Phys. {\bf B207} (1982) 337.\hb
G. Gibbons and K. Maeda, Nucl. Phys. {\bf B298} (1988) 741.}

\lref\GID{S. B. Giddings, Phys. Rev. {\bf D46} (1992) 1347.}

\lref\PRErev{J. Preskill, {\it Do Black Holes Destroy Information?},
Proc. of the International Symposium on Black Holes, Membranes, Wormholes,
and Superstrings, The Woodlands, Texas, 16-18 January, 1992.}

\lref\chgaug{S-W. Chung and S. H. H. Tye, Phys. Rev. {\bf D47} (1993) 4546.}

\lref\eguchi{T. Eguchi, Mod. Phys. Lett. {\bf A7} (1992) 85.}

\lref\HSBW{P. S. Howe and G. Sierra, Phys. Lett. {\bf 144B} (1984) 451.\hb
J. Bagger and E. Witten, Nucl. Phys. {\bf B222} (1983) 1.}

\lref\GSW{M. B. Green, J. H. Schwarz and E. Witten, {\it Superstring Theory},
Cambridge Univ. Press, Vols. 1 and 2, London and New York (1987).}

\lref\KAKU{M. Kaku, {\it Introduction to Superstrings}, Springer-Verlag, Berlin
and New York (1991).}

\lref\LSW{W. Lerche, A. N. Schellekens and N. P. Warner, {\it Lattices and
Strings }, Physics Reports {\bf 177}, Nos. 1 \& 2 (1989) 1, North-Holland,
Amsterdam.}

\lref\confrev{P. Ginsparg and J. L. Gardy in {\it Fields, Strings, and
Critical Phenomena}, 1988 Les Houches School, E. Brezin and J. Zinn-Justin,
eds, Elsevier Science Publ., Amsterdam (1989). \hb
J. Bagger, {\it Basic Conformal Field Theory},
Lectures given at 1988 Banff Summer Inst. on Particle and Fields,
Banff, Canada, Aug. 14-27, 1988, HUTP-89/A006, January 1989. }

\lref\CHAN{S. Chandrasekhar, {\it The Mathematical Theory of Black Holes},
Oxford University Press, 1983.}

\lref\KOULU{C. Kounnas and D. L\"ust, Phys. Lett. {\bf 289B} (1992) 56.}

\lref\PERRY{M. J. Perry and E. Teo, ``Non-singularity of the Exact two
Dimensional String Black Hole'', DAMTP-R-93-1. \hb
P. Yi, ``Nonsingular 2d Black Holes and Classical String Backgrounds'',
CALT-68-1852. }

\lref\GiKi{A. Giveon and E. Kiritsis, ``Axial Vector Duality Symmetry and
Topology Change in String Theory'', CERN-TH-6816-93.}

\lref\kar{S. K. Kar and A. Kumar, Phys. Lett. {\bf 291B} (1992) 246.}
\lref\sfts{K. Sfetsos and A. A. Tseytlin, preprint CERN-TH.6969, to appear.}


\newsec{ Introduction }

There are numerous solutions to string theory, all corresponding to
some conformal field theory.
One wants to use such solutions for phenomenological considerations.
The traditional and older approach was to consider string solutions in
manifolds of the
type $M^4 \times K$, where $M^4$ denotes the four dimensional
Minkowski space-time represented by a free field theory with central charge
$c=4$ (or $\hat c=4$)
and $K$ denotes some internal space represented by a conformal field theory of
appropriate central charge.
Therefore in such constructions only the
internal part requires a non-trivial conformal field theory and
strings propagate in {\it flat} space-times.

However, to describe the very early Universe with a
string inspired cosmological model or to use string theory to shed
light into the singularities of general relativity a non-trivial
conformal field theory is required to describe the {\it curved} space-time the
strings propagate on. In a bosonic string theory it is very
interesting to study the massless excitations of the string, namely
the graviton, the axion and the massless scalar because they govern
the geometry of space-time. In a two dimensional $\s$-model they are
represented by a symmetric tensor (the metric), its antisymmetric
partner (the axion), and the dilaton (the scalar). These fields obey a
set of generalized Einstein's equations, the beta functions equations
(for instance see \CALLAN), which follow by requiring that the
$\s$-model is conformally invariant. The older avenue \GM\ that was
followed to solve these equations was to specialize to the solutions
that have special symmetries, hoping that the equations will become
solvable. The main problem of this approach is that the general form
of the beta function equations is unknown and it is determined order
by order in perturbation theory. Therefore with the old approach it is
very hard to find the exact solution to all orders in perturbation
theory, except in some very special circumstances,
i.e. for instance plane fronted solutions \SANTA.

In order to resolve these problems exact conformal field theories in the
form of coset models based on non-compact groups,
i.e. $G_{-k}/H_{-k}$, $k$ being the central extension of the current algebra,
were introduced in \BN\ as exact string theories.
Based on the equivalence of the
cosets models with the gauged WZW models \GWZW\SCH\ the authors of \BN\
argued that after integrating out the gauge fields a $\s$-model arises
in $dim(G/H)$ space-time dimensions.
The signature of the resulting space-time is intimately connected
to the properties of the group $G$ and the subgroup $H$.
In \BCR\ the coset model $SU(2)_k/U(1)$ was considered in a $\s$-model
approach to classical parafermions, and in \WIT\
the coset model $SL(2,\IR)_{-k}/\IR$ was found to have, in the
semi-classical limit ($k\to \infty$), a two dimensional
black hole interpretation. The latter discovery led to a flurry of activity
and many other solutions were found,
corresponding to black holes, black strings, and other intricate
gravitational singularities, or corresponding to cosmological solutions
\HOHO\BSthree\BShet\groups\GIN\KOULU\NAWIT.
All of these solutions satisfy the perturbative equations for conformal
invariance up to one loop in perturbation theory (to the leading order in
the $1/k$ expansion). However, the interest in these
models stems from the fact that conformal invariance is an exact symmetry and
therefore there must be a way to compute the fields in the $\s$-model to all
orders in the expansion parameter $1/k$. For the metric and dilaton fields
this was achieved for the coset $SL(2,\IR)_{-k}/\IR$ in \DVV, and for
the general gauged WZW model in \BSexa, by following an algebraic Hamiltonian
approach to gauged WZW models. With this method the background fields
in the $\s$-model corresponding to many abelian and
non-abelian cosets were explicitly computed \DVV\BSexa\SFET\BSslsu,
thus providing the first examples of string theory solutions with non-trivial
dependence in the expansion parameter.
Using the Hamiltonian approach type-II supersymmetric and heterotic
string models in curved space-time have also been discussed \BSexa\BSslsu\
(For the type-II superstring based on $SL(2,\IR)/\IR$ see also \JJP).
It has been shown that the background corresponding to the simplest case
$SL(2,\IR)_{-k}/\IR$ verifies the beta function equations in string
perturbation theory up to three \TSEY\ and four \JJP\ loops
(five loops in the type-II supersymmetric case \JJP), up to field
redefinitions.

To compute the axion with the Hamiltonian method seems difficult.
To overcome this
problem and to understand the exact results obtained with the Hamiltonian
method from a field theoretical point of view,
an effective action for gauged WZW models which incorporates all the
quantum effects in the $\s$-model was recently suggested \TSEYT\BSeaction.
Using this effective action the exact metric, antisymmetric tensor and
dilaton fields in the general $\s$-model were obtained in \BSeaction.
The exact metric and dilaton fields one derives
are identical \TSEYT\BSeaction\TSEYTt\
to those obtained with the Hamiltonian method.
However, for the antisymmetric tensor the results are totally new \BSeaction.
In particular, explicit expressions for the cases of the three dimensional
$SL(2,\IR)\otimes \IR/\IR$ (black string) and $SO(2,2)/SO(2,1)$ coset models
were given in \BSeaction.

In this paper we consider another class of exactly conformal models, the so
called {\it chiral } gauged WZW (CGWZW) models which were introduced
in \chgaug. These models have similarities but also major
differences with the usual gauged WZW models which will be pointed out in the
appropriate places. This paper is organized as follows: In section 2 we
discuss general properties of the CGWZW models and we obtain the
semi-classical expressions of the various fields in the $\s$-model. We also
explain in detail how to solve the particle geodesic equations for the
general semi-classical metric.
In section 3 we present an effective quantum action for CGWZW models
by making correspondence with the analogous situation for gauged WZW models.
Using this action the exact expressions for the various fields in the
$\s$-model and the particle geodesics in the exact metric are
obtained. In section 4 we apply the mechanism developed in the previous
sections to the case of the three dimensional model
with $G=SL(2,\IR)$ and $H=SO(1,1)$ (It will be explained shortly why this is a
three dimensional and {\it not} a two dimensional model).
We find that it has certain similarities
with the three dimensional black string.
Finally, we end this paper in section 5 with concluding remarks and discussion.


\newsec{Chiral gauged WZW models }

The two dimensional action for the general CGWZW model is \chgaug

\eqn\chwzw{ S_{cgwzw}=-k\ I_0(g)-k\ I_1(g,A_+,A_-)\ ,
}

\no
where $I_0(g)$ is the WZW action \Wwzw\ defined on a group manifold $G$

\eqn\wzw{I_0(g) ={1\over 8\pi}\int_M d^2\s\
Tr(\pap g^{-1}\pam g)+{1\over 24\pi} \int_B Tr(g^{-1}dg)^3 \ ,
}

\no
and

\eqn\gauge{I_1(g,A_+,A_-)={1\over 4\pi}\int_M d^2\s\ Tr(A_- \pap g g^{-1}
-A_+ g^{-1}\pam g +A_-gA_+g^{-1})\ .
}

\no
In the above $g(\s^+,\s^-)$ is a group element of $G$, and $A_{\pm}(\s^+,\s^-)$
are the gauge fields associated with the subgroup $H$ of $G$.
The gauge fields $A_+$ and $A_-$ may belong to two different subgroups of $G$,
but for simplicity we will only consider the case where the subgroup
is the same. It is important to notice that a term of the type
$Tr(A_+ A_-)$ is absent in \gauge\ in contrast with the
cases of vectorially (and axially)
gauged WZW models \GWZW\SCH\ or deformed gauged
WZW models \BSthree, which we will collectively call gauged WZW models.

The action \chwzw\ is invariant under the following gauge-type transformations

 \eqn\gautr{ g\to \Lambda_- ^{-1}g\Lambda_+\ ,\quad A_{\pm}\to
\Lambda_{\pm}^{- 1}(A_{\pm} -\partial_{\pm})\Lambda_{\pm}\ ,
}

\no
where $\Lambda_{\pm}=\Lambda_{\pm}(\s^{\pm})\in H$.
We see that the parameters of the gauge transformation are functions
of only $\s^+$ or $\s^-$.
\foot{This is a reason for the terminology chiral
gauged WZW models. Also notice that $I_0(g)$ by itself is {\it not} invariant
under the transformation $g\to \L_-^{-1} g \L_+$. For this to be true
$\L_+$ ($\L_-$) must be a function of $\s^-$ ($\s^+$),
i.e. $\L_{\pm}=\L_{\pm}(\s^{\mp})$.}
Consequently this gauge symmetry cannot be
used to remove degrees of freedom, by the usual procedure of gauge fixing.
This is again in contrast with the case of gauged WZW models
where one gauge fixes $dim(H)$ variables due a gauge symmetry of the
type \gautr, but with gauge transformation parameters
which are functions of both variables $\s^+$ and $\s^-$.

The classical equations of motion for the CGWZW
action \chwzw\ follow by varying the $A_-$, $A_+$ and $g$. One
then obtains the following equations

\eqn\eqgen{\eqalign{
&(D_+ g g^{-1})_{H}=0\ ,\qq \cr &(g^{-1}D_- g)_{H}=0\
,\qq \cr &{}\cr}
\eqalign{
&\bl(D^L_-(D_+ g g^{-1})\br)_{G/H}=0\cr &\bl(D^R_+(g^{-1} D_+ g )\br)_{G/H}
=0\cr &\pam A_+=\pap A_-=0\ ,\cr}
}

\no
where the subscripts $H$, $G/H$ imply a projection to the
$H$-subspace or the $G/H$-subspace. The covariant
derivatives are defined as

\eqn\cov{\eqalign{&D_+ g=\pap g +g A_+\ ,\qq D_- g=\pam g -A_- g \cr
&D^R_+=\pap -[A_+,\ \ ]\ ,\qq D^L_-=\pam -[A_-,\ \ ]\ . \cr}
}

In order to obtain the $\s$-model action from the action \chwzw\ one needs to
integrate out the gauge fields $A_{\pm}$.
This integration is easy to perform since the gauge fields appear mostly
quadratically in the action \chwzw, and the corresponding measure in the
functional path integral is just the flat measure $dA_+ dA_-$.
To eliminate the gauge fields through the equations of motion \eqgen,
one has to solve them for $A_{\pm}$.
To do that and for further convenience it is useful to introduce a set
of matrices $\{t_A\}$ in the Lie algebra of $G$
which obey $Tr(t_A t_B)=\eta_{AB}$, $[t_A,t_B]=i f_{AB}{}^C t_C$,
where $\eta_{AB}$ is the Killing metric and $f_{AB}{}^C$ are the structure
constants of the Lie algebra of $G$. The subset of matrices
belonging to the Lie algebra of the subgroup $H$ will be denoted by $\{t_a\}$
with lower case subscripts or superscripts. Then we define the following
quantities

\eqn\difin{\eqalign{&L^H_{\pm}=(g^{-1}\papm g)_H\ , \qq \cr
&R^H_{\pm}=(-\papm g g^{-1})_H\ , \qq \cr
&C_{ab}=Tr(t_a g t_b g^{-1})\ , \cr}
\eqalign{&L^A_{\m}\papm X^{\m}=Tr(g^{-1}\papm g t^A)\cr
&R^A_{\m}\papm X^{\m}=-Tr(\papm g g^{-1} t^A)\cr
& \cr}
}

\no
where $X^{\m},\ \m=0,1,\dots ,d-1$ are $d=dim(G)$ independent parameters in
$g(\s^+,\s^-)\in G$ which will
become the string coordinates in the $\s$-model.
In contrast the $\s$-model arising from the corresponding gauged WZW model
would have $dim(G/H)$ string coordinates because in that case the gauge
symmetry enables us to gauge fix $dim(H)$ parameters in $g(\s^+,\s^-)$.
The solution of the classical
equations of motion \eqgen\ for $A^a_{\pm}\equiv Tr(t^a A_{\pm})$ is

\eqn\lysis{A^a_+=(C^{-1})^a{}_b R^b_{\m}\pap X^{\m}\ ,\qq
A^a_-=L^b_{\m}(C^{-1})_b{}^a \pam X^{\m}\ .
}

\no
Substitution of these expressions back into \chwzw\
gives (to the leading order in the $1/k$ expansion)
the following $\s$-model type of action

\eqn\sm{S_{\s}={k\over 8\pi}\int d^2\s\
(G_{\m\n}+B_{\m\n})\pam X^{\m}\pap X^{\n}\ ,
}

\no
where the explicit forms for the semi-classical metric $G_{\m\n}$ and the
semi-classical antisymmetric tensor (axion) $B_{\m\n}$ are

\eqn\metax{\eqalign{
&G_{\m\n}=g_{\m\n}+(C^{-1})_{ab}L^a_{\{ \m}R^b_{\n \}}\cr
&B_{\m\n}=b_{\m\n}+(C^{-1})_{ab}L^a_{[\m}R^b_{\n]}\ .\cr}
}

\no
The brackets denote symmetrization (when curly) or antisymmetrization
of the appropriate indices.
The $g_{\m\n}$ and $b_{\m\n}$ are the parts of the metric and
axion due to the kinetic and Wess-Zumino terms in $I_0(g)$ respectively.
It can easily be shown that

\eqn\smetric{g_{\m\n}=
L^A_{\m}L^B_{\n}\eta_{AB}=R^A_{\m}R^B_{\n}\eta_{AB}\ ,
}

\no
and that $b_{\m\n}$ are the components of a 2-form defined through the
relation $h=-{3\over 2}db$, where the components of the 3-form $h$ are

\eqn\saxion{h_{\m\n\r}={1\over 2} f_{ABC} L^A_{\m} L^B_{\n} L^C_{\r}
=-{1\over 2} f_{ABC} R^A_{\m} R^B_{\n} R^C_{\r}\ .
}

\no
However, the most efficient way to compute $g_{\m\n}$ and $b_{\m\n}$ is
to use the Polyakov-Wiegman formula \POLWIG\ repeatedly
untill the Wess-Zumino term in $I_{0}(g)$ vanishes identically.
In order to preserve conformal invariance in the $\s$-model approach we need
to take into account the dilaton field \FRATSE\ and to satisfy the perturbative
beta function equations \CALLAN. Up to one loop in perturbation theory
the dilaton can be identified as the finite part of the determinant of the
matrix we obtain by integrating out the gauge fields
\BSthree\KIR\ (see also \TSESC\ for further clarifications). Therefore
in our case the semi-classical $(k\to \infty)$ expression for the dilaton is

\eqn\detd{\Phi(X)=\ln ({\rm det} C) +{\rm const.}\ .
}

\no
Consequently the total $\s$-model action which is conformally invariant
up to one loop in perturbation theory takes the form

\eqn\tsmg{S=S_{\s}-{1\over 8\pi}\int_M d^2\s\ \sqrt{\g} R^{(2)}(\g)\Phi(X)\ ,
}

\no
where $\g$, $R^{(2)}(\g)$ are the determinant of the world-sheet metric and
the world-sheet curvature respectively.
The infitite part of the determinant $\det(C)$,
combines with the Haar measure for the group $G$
(which together with the flat measure $dA_+ dA_-$ for the gauge fields
provides the correct measure in the path integral for the action \chwzw)
to give the correct measure for the
$\s$-model which is none other but $\sqrt{-G}$, where $G=\det(G_{\m\n})$.
Namely the following relation must be true

\eqn\theo{({\rm Haar})/{\rm det}(C)=\sqrt{-G}\ .
}

\no
Furthermore because of the identification \detd\ one can rewrite \theo\ as

\eqn\theorem{ e^\Phi \sqrt{-G} = \ {\rm (Haar)}\ .
}

\no
One now notices that the right hand side of the previous relation is
purely group theoretical and therefore one expects that \theorem\ will be
true even when we include all the $1/k$ corrections. Namely, although
$G_{\m\n}$ and $\Phi$ would be nontrivial functions of $k$ the following
combination would remain $k$-independent

\eqn\theoremm{ e^\Phi \sqrt{-G}\ ({\rm any}\ k)\ =\ e^\Phi \sqrt{-G}\ ({\rm
at}\ k=\infty )\ .
}

\no
We will prove \theorem\ and \theoremm\ for any CGWZW model
at the end of section 3.
Similar relations to \theorem, \theoremm\
hold for the $\s$-model arising from a general gauged WZW model as well.
These were conjectured in \KIR\ for the abelian $SL(2,\IR)_{-k}/\IR$ coset
case,
and in \BSthree\ where they were generally formulated for any gauged WZW model,
by using the path integral measure argument that led to \theorem, \theoremm.
Subsequently their validity was explicitly checked for many abelian and
non-abelian cases in \BShet\BSexa\SFET\BSslsu\ and proved generally for any
gauged WZW model in \TSEYTt.

As we have already mentioned the string coordinates $X^{\m}$
in the $d$-dimensional
$\s$-model arising form a general CGWZW model are functions of
the $d=dim(G)$ parameters in the group element $g(\s^+,\s^-) \in G$.
Therefore the ranges in which they take values are completely determined by
the group theory for $G$ and in that sense the $X^{\m}$'s are global
coordinates.
In the case of gauged WZW models finding global coordinates was a rather
delicate procedure. In that case the corresponding $\s$-model depends on
$dim(G/H)$ string coordinates which are the $dim(G/H)$ independent
$H$-invariant combinations one can form out of
the $dim(G)$ parameters in $g(\s^+,\s^-)$. The necessary techniques for finding
the global space in this case were developed and applied in various
examples in \BSglo\BSexa\BSslsu.

Having global coordinates is sometimes not sufficient to get a feeling
for the geometry;
one also needs to know the behavior of the particle geodesics by solving
the usual geodesic equations
$\ddot{X}^{\m}+\G^{\m}{}_{\n\r}\dot{X}^{\n}\dot{X}^{\r}=0$.
However, these equations may seem completely unmanageable if the metric
that emerges from \metax\ is complicated.
To get around this problem we first, as in \BSglo, dimensionally reduce the
CGWZW action \chwzw\ by taking all fields to be functions of only
$\tau$, instead of $\tau$ and $\s$. This corresponds to a string shrunk to a
point particle. Therefore the action we consider is

\eqn\action{S={-k\over 4\pi}\int d\tau \
Tr({1\over 2}\pat g^{-1}\pat g+a_-\pat g g^{-1}-a_+g^{-1}\pat g
+a_-ga_+g^{-1} )\ ,
}

\no
where $g(\tau)\in G$ is a group element and $a_{\pm}(\tau)$ are two gauge
potentials in the Lie algebra of $H$. Two gauge potentials are needed for our
purposes. The model is invariant for rigid ($\tau$ and $\s$ independent)
gauge transformations of the form \gautr. This was expected since already
for the full two dimensional action \chwzw\ the parameters of the
gauge transformation were functions of only $\s^+$ or only $\s^-$.
Consider the equations of motion

\eqn\eqs{ \bl(g^{-1}D_-g \br)_H = 0 = \bl(D_+gg^{-1} \br)_H\ ,
\quad D^R_+(g^{-1}D_-g)=0\ ,
\quad \dot{a}_+=\dot {a}_-=0\ ,
}

\no
where we have defined the ``covariant'' derivatives on the worldline
$D_+ g=\dot{g} + g a_+$, $D_- g=\dot g -a_- g$, $D^R_+=\pat -[a_+,\ \ ]$.
The solution of these equations for $g(\tau)$ will provide the required
geodesics by virtue of the fact that $g(\tau)$ contains
all particle coordinates $X^{\m}(\tau)$.
{}From \eqs\ one can see that $a_{\pm}(\tau)=\a_{\pm}$, where $\a_{\pm}$ are
two constant matrices in the Lie algebra of $H$. The first and third equation
yield the equation $\dot p=[\a_+,p]$,
where $p(\tau)\equiv (g^{-1}D_- g)_{G/H}$,
whose solution is $p(\tau)\equiv \exp(\a_+ \tau) p_0 \exp(-\a_+ \tau)$, with
$p_0$ a constant matrix in the Lie algebra of $G/H$. From the definition of
the matrix $p(\tau)$ and the first equation one determines $g(\tau)$ to be

\eqn\sol{ g(\tau)=\exp{(\a_-\ \tau)}\ g_0\ \exp{\bl((\a_+ +p_0)\ \tau\br)}\
\exp{(-\a_+\ \tau)}\ ,
}

\no
where $g_0$ is a constant group element that characterizes the initial
conditions. Finally, replacing this form into the remaining second equation in
\eqs\ yields a constraint among the constants of integration which completely
determines the constant matrix $\a_-$ in terms of the constant matrices
$g_0$, $a_+$ and $p_0$

\eqn\const{  \a_- =- \bl(g_0(\a_+ +p_0) g_0^{-1} \br)_H\ .
}

\no
The number of independent parameters in \sol\ is:
$dim(G)$ parameters from $g_0$, plus
$dim(H)$ parameters from $\a_+$, plus $dim(G/H)$ parameters from $p_0$, giving
a total number of $2 dim(G)$ parameters. This is
precisely the number of initial positions $X^{\m}(0)$ and velocities
$\dot X^{\m}(0)$ needed for the general geodesic in $dim(G)$ dimensions.
Therefore \sol, with the condition \const, is the general geodesic solution.
The Lagrangian defined in \action\ is rewritten
in terms of the line element
$(ds/d\tau)^2={k\over 8\pi}\dot{X}^{\m}\dot{X}^{\n} G_{\m\n}(X)$,
because all the other string modes drop out in the point particle limit.
Therefore, if we substitute the solution \sol\ in the
Lagrangian defined in \action\ we find the value of $(ds/d\tau)^2$ for the
geodesic solution. This gives

\eqn\dsds{ \bl({ds\over d\tau }\br)^2= {k\over 8\pi} Tr(p_0^2-\a_-^2)\ .
}

\no
Now by choosing the various constant matrices we have control on whether the
geodesic is light-like, time-like or space-like.

The model with $G=SL(2,\IR)$ and $H=SO(1,1)$ is perhaps the simplest model
one may consider. We will discuss it in
section 4 where we will derive the corresponding
conformally exact metric,
antisymmetric tensor and dilaton fields using the methods of section 3.

\newsec{The effective action for chiral gauged WZW models}

The effective quantum action for any field theory is derived by introducing
sources and then applying a Legendre transform \ITZ.
The effective action, which is then used as a classical field theory,
incorporates all the higher loop effects. In this section we suggest such an
effective action for the CGWZW models. The main idea we follow
was developed for gauged WZW models in \TSEYT\BSeaction.

It is useful to make a change of variables for the gauge fields
$A_{\pm}=\partial_{\pm}h_{\pm}h_{\pm}^{-1}$, where $h_{\pm}(\s^+,\s^-)\in H$.
After picking up a
determinant and an anomaly from the measure, the path integral
is rewritten with a new form for the action \POLWIG\SCH\chgaug

\eqn\agroup{ S_{cgwzw}=-kI_0(h_-^{-1}gh_+)
+(k-2g_H)I_0(h_-^{-1})+(k-2g_H)I_0(h_+)\ ,
}

\no
which is gauge invariant under $g\to \Lambda_- ^{-1} g \Lambda_+$ and
$h_\pm\to \Lambda_{\pm}^{-1} h_\pm$, where
$\Lambda_{\pm}=\Lambda_{\pm}(\s^{\pm})$ as in \gautr.
The new path integral measure is the
Haar group measure  $Dg\ Dh_+\ Dh_-$. The action \agroup\ is
similar to the classical WZW action \wzw: the first term is
appropriate for the group $G$ with central extension $(-k)$, and
the second and third terms are
appropriate for the subgroup $H$ with central extension $(k-2g_H)$.
Defining the new fields
$g'=h_-^{-1}gh_+$, $h'=h_- ^{-1}$, $h''=h_+$ and taking advantage of the
properties of the Haar measure, we can rewrite the measure and action in
decoupled form $Dg'\ Dh'\ Dh''$ and $S=-kI_0(g')+(k-2g_H)(I_0(h')+I_0(h'')$.
This decoupled form emphasizes the close connection to the WZW path integral,
and gives us a clue for how to guess the effective quantum action.

However, $g',h',h''$ are not really decoupled, since we must consider sources
coupled to the {\it original fields}. Indeed, to derive the quantum effective
action one must introduce source terms and perform a Legendre transformation.
Since these coupled $g',h',h''$ integrations are not easy to perform, we will
introduce, as in the case of gauged WZW models \BST,
sources only for the gauge invariant combinations $g',h',h''$.
Since for each one of these fields the action is that of a WZW model the effect
is a shift in $k$ (as it was argued in \TSEYT\ based on the perturbative
analysis of \SHIF\SHIFM), which however is different in the various terms in
\agroup.
For the first term, $(-k)\to (-k+g_G)$, and for the second and third term
$(k-2g_H)\to (k-2g_H)+g_H=k-g_H$, where $g_G$, $g_H$ are the dual Coxeter
numbers for the group $G$ and the subgroup $H$.
Therefore the effective action for the CGWZW models we suggest, is

\eqn\egwzw{ S^{eff}_{cgwzw}
=(-k+g_G)I_0(h_-^{-1}gh_+)-(-k+g_H)I_0(h_-^{-1})-(-k+g_H)I_0(h_+) \ .
}

\no
The similar effective
action for the vectorially gauged WZW models can be obtained by
omitting the third term and replacing $h_-^{-1}$ by $h_-^{-1} h_+$ in
the second term \BST\TSEYT\BSeaction\
(For axial gauged WZW models see \BSeaction).
We should point out that in \egwzw\ we have neglected possible field
renormalizations \nonl\ for the group elements $g,h_{\pm}$ since they give
rise to non-local terms in the $\s$-model action \TSEYTt.
The action \egwzw\ may now be rewritten back in terms of {\it classical}
fields $g,A_+,A_-$
by using the definitions given before and the Polyakov-Wiegman formula \POLWIG.
We obtain

\eqn\effgwzw{ \eqalign {
     & S^{eff}_{cgwzw}=(-k+g_G)\bl [I_0(g)+I_1(g,A_+,A_-)
       +{g_G-g_H\over -k+g_G}I_2(A_+,A_-)\br ] \ , \cr
&I_2(A_+,A_-)=I_0(h_-^{-1})+I_0(h_+)\ . }
}

\no
Note that $I_2(A_+,A_-)$ is gauge invariant.
Our proposed effective action differs from the purely classical action
\chwzw\ by the overall renormalization $(-k+g_G)$ and by the additional term
proportional to $(g_G-g_H)$.
In the large $k$ limit (which is equivalent to small
$\hbar$) the effective quantum action reduces to the classical action, as it
should.
This is not yet the end of the story, because what we are really interested in
is the effective action for the $\s$-model after the gauge fields are
integrated out.
At the outset, with the classical action, the path integral
over $A_\pm$ was purely Gaussian, and therefore it could be performed
by simply
substituting the classical solutions for $A_\pm=A_\pm (g)$ back into the
action. This integration also introduces an anomaly which can be computed
exactly as a one loop effect. The anomaly gives the dilaton piece to be
added to the effective action. In order to obtain the exact
dilaton we need to perform the $A_\pm$ integrals with the effective action,
not the classical one. However, in \effgwzw\ the terms in $I_2(A_+,A_-)$ are
non-local in the $A_\pm$ (although they are local in $h_\pm$).
For instance, $I_0(h_+)\sim \int Tr(A_+\partial_-h_+h_+^{-1})+\dots $, and we
cannot write $\partial_-h_+h_+^{-1}$ as a local function of $A_+$.
Furthermore, if $H$ is non-abelian $I_2(A_+,A_-)$ has additional non-linear
terms. So, if we believe that the quantum effective action is indeed \effgwzw,
then the effective $\s$-model action we are seeking seems to be generally
non-local even in the abelian case. This was also true for the effective
action for gauged WZW models, as it was discussed in \TSEYT\BSeaction.
As in \BSeaction, we can isolate the local contribution to the $\s$-model
by concentrating on the zero mode sector of \effgwzw.
To restrict ourselves to the zero mode sector we employ the same
dimensional reduction technique as before
by taking all the fields as functions of only $\tau$ (i.e. worldline rather
than world-sheet). This extracts the low energy point particle content of the
string and it captures the entire local contribution to the $\s$-model.
The derivatives $\partial_\pm$ get replaced by $\partial_\tau$
and $A_\pm$ get replaced by $a_\pm=\partial_\tau h_\pm h^{-1}_\pm $. Then all
non-local and non-linear terms drop out and we obtain the effective action in
the zero mode sector

\eqn\actionzero{\eqalign{S_{eff}=&{-k+g_G\over 4\pi}\int d\tau \
Tr({1\over 2}\pat g^{-1}\pat g+a_-\pat g g^{-1}-a_+g^{-1}\pat g
+a_-ga_+g^{-1} )\cr
&-{g_G-g_H\over 8\pi} \int d\tau\ Tr(a_+^2 +a_-^2)\ .\cr }
}

\no
This action is gauge invariant for
rigid ($\s^{\pm}$-independent) gauge transformations $\Lambda_{\pm}$.
Most notably the path
integral over $a_\pm$ is now Gaussian, and this permits the elimination of
$a_\pm$ through the classical equations of motion

\eqn\classe{ (D_+gg^{-1})_H=-{g_G-g_H\over k-g_G}\ a_-\ ,\qquad
               (g^{-1}D_-g)_H={g_G-g_H\over k-g_G}\ a_+\ , \
}

\no
with the same ``covariant'' derivatives $D_\pm$ on the worldline as before.
The system of equations \classe\ is linear
and algebraic in $a_{\pm}$ and therefore it can easily be solved.
Its solution for $a_{\pm}$ is

\eqn\sola{\eqalign{&a_+=
\bigl(C^t C-\l^2 I\bigr)^{-1}\bigl(C^t R^H-\l L^H)\bigr)\cr
&a_-=
\bigl(CC^t-\l^2 I\bigr)^{-1}\bigl(C L^H-\l R^H)\bigr)\ ,\cr}
}

\no
where $C^t$ denotes the transpose matrix of $C$ and $\l={g_G-g_H\over k-g_G}$.
The rest of the quantities appearing in \sola\
were defined in \difin\ ($L^H, R^H$ are the point particle analogs of
$L^H_{\pm}, R^H_{\pm}$).
Substitution of these expressions back into \actionzero\ gives

\eqn\apoint{ S^{eff}_{point} = {k-g_G\over 8\pi}
\int d\tau\ G_{\m\n}\pat X^{\m} \pat X^{\n}\ ,
}

\no
where the metric $G_{\m\n}$ is defined as


\eqn\metric{G_{\m\n}=
g_{\m\n}+(\tilde V^{-1} C^t)_{ab}L^a_{\{ \m}R^b_{\n \}}
-\l \tilde V^{-1}_{ab}L^a_{\m}L^b_{\n}
-\l V^{-1}_{ab}R^a_{\m}R^b_{\n}\ ,
}

\no
where $g_{\m\n}$ was defined in \smetric\ and for convenience we have
defined the symmetric matrices

\eqn\vv{V_{ab}=(CC^t-\l^2 I)_{ab}\ , \qq \tilde V_{ab}=(C^t C-\l^2 I)_{ab}\ .
}

\no
In the $k$ large limit the metric \metric\ tends to the corresponding
semi-classical expression in \metax\ as it should.

To obtain the axion $B_{\m\n}$ we need to retain the $\partial_\pm$ on
the worldsheet and then read off the coefficient of
${1\over 2}(\pam X^{\m} \pap X^{\n} - \pam X^{\n} \pap X^{\m} )B_{\m\n}(X)$.
As already explained above we cannot do this fully
because of the non-local terms and non-abelian non-linearities, but we can
still obtain the local contribution to the axion as follows.
We formally replace the $R^H,L^H$
in the expressions for $a_\pm$ and elsewhere by $R^H_\pm,L^H_\pm$,
where $R^H_\pm$ and $L^H_\pm$ were defined in \difin.
We justify this step
by the conformal transformation properties for left and right movers. We then
substitute these forms of $A_\pm$ back into the action \effgwzw\ and extract
the desired metric and axion from the quadratic part.
The expression we find for the metric $G_{\m\n}$ is of course the same
as in \metric, whereas for the axion $B_{\m\n}(X)$ we find the following result


\eqn\axion{B_{\m\n}=b_{\m\n}
+(\tilde V^{-1} C^t)_{ab} L^a_{[\m}R^b_{\n]}\ ,
}

\no
where $b_{\m\n}$ was defined in \saxion. As it was the case with the metric
\metric\ the axion \axion\ tends to the corresponding semi-classical
expression in \metax\ for large $k$.

To obtain the exact dilaton we must compute the anomaly in the integration
over $A_\pm$. However, as it was the case with the metric and the axion,
the local part of the dilaton can be obtained by going to the point
particle limit.
The effective action \actionzero\ contains a quadratic part in the gauge
fields which can be rewritten as

\eqn\actionn{{-k+g_G\over 4\pi}
\int d\tau\ Tr\bigl(a_- C a_+ +{\l \over 2}(a_- ^2 +a_+ ^2) \bigr)\ .
}

\no
Integrating out the gauge fields $a_{\pm}$ gives a determinant that produces
the exact dilaton by identifying, as in section 2,
$e^{\Phi}=({\rm determinant})$. The result we obtain is


\eqn\dilaat{\Phi(X)={1\over 2}\ln \bl( \det (V) \br) +{\rm const.} \ .
}

\no
Again it is easy to see that the above expression for the dilaton tends,
for large $k$, to the semi-classical result \detd.
The expressions for the metric, the antisymmetric tensor and the dilaton in the
$\s$-model arising from a general gauged WZW model were found in \BSeaction.
It is worth pointing out that they can be obtained from the corresponding
expressions in \metric, \axion, \dilaat\ if we formally make the substitution
$C\to C-(\l +1) I$ and, among the $dim(G)$ string coordinates, restrict to the
$dim(G/H)$ combinations that are $H$-invariant.

Let us determine the particle geodesic equations for the exact metric \metric.
So, we seek a solution
to the classical equations of motion given by \classe\ and

\eqn\classs{ D^R_+(g^{-1}D_- g)=-\partial_\tau a_+\ ,
}

\no
which follows from varying $g$, and
where $D^R_+$ was defined just below equation \eqs. The method for solving
these equations is identical to the one that led to equations \sol, \dsds\ and
it will not be repeated. We will only give the solution
which as a function of proper time $\tau$ is

\eqn\solu{\eqalign{
&g(\tau)=\exp \bl ({k-g_G\over k-g_H}\ \a_-\ \tau\br ) \ g_0\ \exp
\bl((\a_+ +p_0)\ \tau \br)\ \exp\bl(-{k-g_G\over k-g_H}\ \a_+\ \tau\br)\ ,\cr
&{\rm with}\qq \a_-=- \bl(g_0(\a_+ +p_0) g_0^{-1} \br)_H \ ,\cr}
}

\no
where $\a_{\pm}$, $p_0$ are constant matrices in the Lie algebra of $H$
and $G/H$
respectively, and $g_0$ is a constant group element in $G$. These matrices,
define the initial conditions for any geodesic at $\tau=0$.
The line element evaluated at this general solution becomes

\eqn\lineel{ \bl ( {ds\over d\tau}\br )^2 = {k-g_G\over 8\pi} \
 Tr \bl (p_0^2 -\a_- ^2 +{\l \over (\l+1)^2}\ \a_- ^2 +{\l \over \l+1}\
\a_+ ^2 \br)\ .
}

\no
For the particular example considered in section 4 we have verified
that \solu, \lineel\ indeed solve the geodesic equations which are obtained
from the exact metric in (4.5) below.

In the rest of this section we prove the theorems \theorem, \theoremm.
Let us denote the inverses of $L^A_{\m}$, $R^A_{\m}$, which were defined in
\difin, by $L^{\m}_A$ and $R^{\m}_A$ respectively.
The following properties are hepful in the algebraic
manipulations needed for the proof

\eqn\prop{\eqalign{&L^A_{\m} L^{\m}_B=\eta^A{}_B\ ,
\quad L^A_{\m} L^{\n}_A=\d^{\m}{}_{\n}\ ,
\quad  R^A_{\m} R^{\m}_B=\eta^A{}_B\ ,
\quad R^A_{\m} R^{\n}_A=\d^{\m}{}_{\n} \cr
&V^{-1}C=C \tilde V^{-1}\ ,\qq \tilde V^{-1} C^t=C^t V^{-1} \ . \cr }
}

\no
Let us rewrite the exact metric \metric\ in the following way

\eqn\metvv{G_{\m\n}=g_{\m\r} \tilde G^{\r}{}_{\n}\ ,
}

\no
where $g_{\m\r}$ was defined in \smetric\ and

\eqn\mettvv{\eqalign{
\tilde G^{\r}{}_{\n}&=\d^{\r}{}_{\n} +(\tilde V^{-1} C^t)_{ab} (L^{a\r}R^b_{\n}
+L^a_{\n} R^{b\r})-\l \tilde V^{-1}_{ab} L^{a\r} L^b_{\n}
-\l V^{-1}_{ab} R^{a\r}R^b_{\n} \cr
&=\d^{\r}{}_{\n} +\tilde C_{a'b'} S^{a'\r} S^{b'}_{\n}\ ,\cr}
}

\no
where $S^{a'}_{\m}=(L^a_{\m},R^a_{\m})$ and the $2 dim(H) \times 2 dim(H)$
dimensional symmetric matrix $(\tilde C_{a'b'})$ is defined as

\eqn\tcc{(\tilde C_{a'b'})=\pmatrix{-\l \tilde V^{-1} & \tilde V^{-1} C^t\cr
C \tilde V^{-1} & -\l V^{-1} \cr}\ .
}

\no
In order to compute $\det(G_{\m\n})$ we need $\det(\tilde G^{\r}{}_{\n})$.
We have

\eqn\detg{\eqalign{\det(\tilde G^{\r}{}_{\n})
&=\det(\d^{\r}{}_{\n} +\tilde C_{a'b'} S^{a'\r} S^{b'}_{\n}) \cr
&=\det(\eta_{a'b'} +\tilde C_{a'c'} S^{c'\m} S_{b'\m}) \cr
&=\det \pmatrix{I- \tilde V^{-1} (C^t C +\l I) & (\l +1) \tilde V^{-1} C^t \cr
(\l +1) V^{-1} C & I -V^{-1} (C C^t +\l I) \cr} \cr
&=(\l+1)^{2 dim(H)}\ \det\pmatrix{-\l \tilde V^{-1} & \tilde V^{-1} C^t\cr
V^{-1} C & -\l V^{-1} \cr} \cr
&=(\l+1)^{2 dim(H)}\ \det\pmatrix{\tilde V^{-1} & 0\cr 0 & V^{-1}\cr}
\det\pmatrix{-\l I & C^t \cr C & -\l I \cr} \cr
&=-(\l+1)^{2 dim(H)}\ {\det(V^{-1})}\ . \cr }
}

\no
The Haar measure for the group $G$ is given by
$\det^{1/2}(g_{\m\r})$. Therefore by using \metvv, \detg, \dilaat\ one easily
establishes the validity of the theorems \theorem, \theoremm\ for any
CGWZW model.


\newsec{ Chiral gauging with $G=SL(2,\IR)$ and $H=SO(1,1)$ }

Let us work out explicitly the details in the simple case where $G=SL(2,\IR)$
and $H=SO(1,1)$. If one considers the corresponding gauged WZW model
one obtains a two dimensional black hole \WIT.
However in our case we will find a three dimensional model which is
related to the three dimensional black string model \HOHO.

It is convenient to parametrize the group element $g\in SL(2,\IR)$ as

\eqn\groupsl{g=\pmatrix{u&a\cr -b&v\cr}\ , \qq ab+uv=1\ .
}

\no
The subgroup generator is $t_0={1\over \sqrt{2}} \s_3$, where $\s_3$ denotes
the usual third Pauli matrix,
and the dual Coxeter numbers are $g_G=2$, $g_H=0$.
Using \difin\ we compute the following quantities necessary for
the evaluation of the various fields in the $\s$-model

\eqn\rlc{\eqalign{&L^0_{\m}=-\sqrt{2}\pmatrix{b\cr 0\cr u}\ ,\qq
R^0_{\m}=-\sqrt{2}\pmatrix{b\cr v\cr 0} \cr
&C_{00}=uv-ab\ ,\qq {\rm where}\quad X^{\m}=(a,u,v)\ .}
}

\no
Then using \metric, \axion, \dilaat\ we find the following expression for the
line element, the antisymmetric tensor and the dilaton

\eqn\ds{\eqalign{
&ds^2={k/2 \over uv-ab+\l}\bl({b\over a}\ da +{1\over a}\ d(uv)\br)\ da
+{k/2 \over uv-ab-\l}\ dudv -{\bl(d(uv)\br)^2 \over (uv-ab)^2 -\l^2} \cr
&B=\ln a\ du \wedge dv +2 {uv-ab\over (uv-ab)^2 -\l^2}
 (bv\ da \wedge du -bu\ da \wedge dv -uv\ du \wedge dv) \cr
& \Phi={1\over 2} \ln\bl((uv-ab)^2-\l^2\br) +{\rm const.}\ .}
}

\no
Next we make the following change of variables

\eqn\uva{uv={1\over 2}(1+\l-r)\ ,\quad {u\over v}=e^{2t}\ ,\quad
a=e^x \bl({1\over 2}(1-\l +r)\br)^{1/2}\ .
}

\no
In this set of variables the various fields in \ds\ take the form
(after rescaling $t$, $x$ and dropping out total derivative terms in the
expression for $B$)

\eqn\tdbs{\eqalign{
&ds^2=-(1-{1+\l \over r})\ dt^2 -(1+{1+\l \over r-2\l})\ dx^2
+{1\over 4\l}\ {dr^2 \over (r-\l)^2-1}  \cr
&B=(1-\l)\ {r-\l\over (r-\l)^2-\l^2}\ dt \wedge dx \cr
&\Phi={1\over 2}\ln\bl(r(r-2\l)\br) +{\rm const.}\ .\cr}
}

\no
The semi-classical limit ($k\to \infty$, $\l\to 0$) of the above expressions
was previously obtained in \kar.
It is possible to relate the above results to the corresponding
exact results for the three dimensional black string model \HOHO\ one
obtains by axially gauging the subgroup $H=SO(1,1)$ of the direct product
group $G=SL(2,\IR) \otimes SO(1,1)$.
The exact metric and dilaton for this model
have been found in \SFET\ and the exact antisymmetric tensor in \BSeaction

\eqn\emda{\eqalign{
&ds^2=-(1-{r_+ \over r})\ dt^2 +(1-{r_- -r_q \over r-r_q})\ dx^2
+{1\over 4\l}\ {dr^2 \over (r-r_+)(r-r_-)} \cr
&B=\sqrt{{r_- -r_q \over r_+}}\ {r-r_+ \over r-r_q}\ dt \wedge dx \cr
&\Phi={1\over 2}\ln\br(r(r-r_q)\br) +{\rm const.}\ ,\cr }
}

\no
where $r_+=(\r^2 +1)\ C'$, $r_-=(\r^2 +2/k)\ C'$, $r_q=2/k\ C'$.
The embedding of
the subgroup $H$ into the group $G$ is parametrized by the positive parameter
$\r^2$ and $C'$ is a constant. If $C'=2(\l+1)$, $\r^2=-1/2$ we get
$r_+=\l+1$, $r_-=\l-1$ and $r_q=2\l$. The fact that $\r^2 <0$ means that $x$
is space-like and therefore we should also
analytically continue $x\to i x$. Then one can verify that the expressions for
the exact metric and dilaton of the black string \emda\
become the corresponding expressions of our model in \tdbs.
However, the expression one obtains for the antisymmetric tensor is
$B={\l +1-r \over r-2 \l}\ dt \wedge dx$ which differs from the one in \tdbs.
There is agrement only in the semi-classical limit ($k\to \infty$, $\l \to 0$)
(up to a constant piece).
In the semi-classical limit the above correspondence between the two models
was first observed in \kar.



\newsec{ Conclusion }

We have suggested an effective action for CGWZW models,
by making contact with the analogous problem for gauged WZW models.
The effective action for the latter reproduced correctly the exact geometry
derived before in the Hamiltonian formalism and this is essentially the
justification of our approach.
\foot{In the case of CGWZW models the
metric \metric\ and dilaton \dilaat\ can also be derived via the Hamiltonian
formalism by using the Hamiltonian

$${\cal H}={J^2_G\over k-g_G}-2 {J^2_H\over k-g_H}
+{\bar J^2_G\over k-g_G}-2 {\bar J^2_H\over k-g_H}\ ,$$

\no
where $J_G$, $J_H$ belong in the Lie algebras
of $G$ and $H$ respectively and act as first order differential operators
in the group parameter space (similarly for $\bar J_G$, $\bar J_H$). }

Using the effective action we derived the exact expressions for the metric,
the antisymmetric tensor and the dilaton
fields in the $\s$-model arising from the general CGWZW model.
We explicitly considered the three dimensional case with $G=SL(2,\IR)$
and $H=SO(1,1)$ and saw that it is related to the three dimensional
black string which however arises in a different context,
i.e. by axially gauging the four dimensional
direct product group $SL(2,\IR)\otimes \IR$ by a total translation.
The correspondence is exact for the metric and dilaton but only semi-classical
for the antisymmetric tensor.

We think that it would be interesting to extend this
construction to include the supersymmetric case,
along the lines of \WITnm\BShet\TSEYTt\
and to investigate higher dimensional models in the context of CGWZW models.


\bs\bs

\centerline { Note added }

\bs

The apparent disagreement of the expressions for the exact antisymmetric
tensors in the cases of $SL(2,\IR) \otimes \IR/\IR$ gauged WZW model
and $SL(2,\IR)/\IR$ chiral gauged WZW model is resolved in \sfts.
The two backgrounds are related by local field redefinitions.

\listrefs
\end